\documentclass[a4paper]{jpconf}
\usepackage{graphicx}
\begin{document}
\title{A molecular dynamics and circular dichroism study of a novel synthetic antimicrobial peptide}

\author{N P Rodina, A N Yudenko, I N Terterov and I E Eliseev}

\address{Laboratory of nanobiotechnologies, St.~Petersburg Academic university, Khlopin str., 8/3, St.~Petersburg 194021, Russian Federation}

\ead{eliseev@spbau.ru}

\begin{abstract}
Antimicrobial peptides are a class of small, usually positively charged amphiphilic peptides that are used by the innate immune system to combat bacterial infection in multicellular eukaryotes. Antimicrobial peptides are known for their broad-spectrum antimicrobial activity and thus can be used as a basis for a development of new antibiotics against multidrug-resistant bacteria. The most challengeous task on the way to a therapeutic use of antimicrobial peptides is a rational design of new peptides with enhanced activity and reduced toxicity. Here we report a molecular dynamics and circular dichroism study of a novel synthetic antimicrobial peptide D51. This peptide was earlier designed by Loose et al. using a linguistic model of natural antimicrobial peptides. Molecular dynamics simulation of the peptide folding in explicit solvent shows fast formation of two antiparallel beta strands connected by a beta-turn that is confirmed by circular dichroism measurements. Obtained from simulation amphipatic conformation of the peptide is analysed and possible mechanism of it's interaction with bacterial membranes together with ways to enhance it's antibacterial activity are suggested. 
\end{abstract}

\section{Introduction}
Antimicrobial peptides (AMPs) are generally defined as short peptides of less than 50 residues with overall positive charge and substantial portion of hydrophobic residues. Natural antimicrobial peptides found in different multicellular eukaryotes and are an important part of the innate immune system that defend host organism against pathogenic microbes\cite{amp-book}. Broad-spectrum activity of antimicrobial peptides against multidrug-resistant bacteria make them potential next-generation antibiotics\cite{review}. 

Currently more than 2000 AMP sequences can be found in Antimicrobial Peptide Database\cite{database}. Most of the studied antimicrobial peptides with known structure belong to three major structural classes: $\alpha$-helix, $\beta$-sheet or combined $\alpha$+$\beta$-structures. Spatial structure (conformation) of a peptide molecule defines a specific distribution of physico-chemical properties such as an electric charge and hydrophobicity on a molecular surface. Presence of a certain amphipatic conformation is essential for an antimicrobial activity of a peptide. It is widely accepted that antimicrobial peptides target the lipid bilayer of the bacterial cell membrane, and their selectivity is based on higher electrostatic affinity of cationic peptides to negatively charged bacterial membranes than to electrically neutral lipids forming membranes of mammalian cells\cite{structure-activity}.

Most of the natural antimicrobial peptides are not enough active to be used as therapeutic agents straightforwardly. Thus, new design approaches are required to build synthetic antimicrobial peptides with enhanced activity. Among other modern methods of peptide design there is a method of \emph{single amino acid replacement} in natural antimicrobial peptide sequences, method of \emph{artificial neural network} that had been trained for AMP recognition and ranks a large set of random generated sequences and \emph{linguistic model} where new antimicrobial peptides are generated based on grammatical rules derived from naturally occurring AMPs\cite{lingvo}. 

Despite the mentioned approaches are rather good for generation of quazirandom peptide sequences that exhibit antimicrobial activity, they do not explicitly take into account the structure of antimicrobial peptides, that is crucial for their activity and can be used in further improvement of their properties. Here we use molecular dynamics simulation in order to complete linguistic model with structural information and give insights into molecular basis of an activity of the designed synthetic peptide D51 (FLFRVASKVFPALIGKFKKK)\cite{lingvo}. Structure obtained by molecular dynamics simulation is used to build hypothesis about peptide interaction with bacterial membranes and propose modifications to enhance it's activity.

\section{Materials and methods}
\subsection{Molecular dynamics simulation}
Two MD simulations of the peptide in an explicit solvent were performed using GROMACS 4.5.4 software package\cite{gromacs}. The solvent was modeled using the simple-point charge (SPC) water model. The peptide and peptide-water interactions were described by the GROMOS 43a2 united-atom force field. Simulations were performed using periodic boundary conditions and a time step of 2 fs was used. The temperature was weakly coupled to $T=300$ K with time constant $\tau_T=0.5$ ps using the Nose-Hoover thermostat. The pressure was weakly coupled to 1 atmosphere with time constant $\tau_P=2$ ps using Parrinello-Rahman barostat.

Initial linear structure of the peptide for the first simulation was build with dihedral angles $\phi=\psi=180^{\circ}$ for each residue. Final structure of the peptide after the first simulation was used as the starting structure for the second simulation. DSSP\cite{dssp} algorithm was used for the secondary structure assignment. PLATINUM program\cite{platinum} was used for the calculation of molecular hydrophobicity potential and PyMol package was used for visualization.

\subsection{Peptide preparation}
The peptide D51 (FLFRVASKVFPALIGKFKKK-OH, $MW_{theor.}=2324.96\;Da$) was synthesized using Fmoc (fluorenylmethoxycarbonyl) chemistry with Symphony synthesizer at the Peptide Technologies, Inc. and was supplied as dry powder. The correct mass $MW_{exp.}=2325.2\;Da$ was observed using Voyager-DE STR MALDI-TOF mass spectrometer (Applied Biosystems) and the purity of the peptide was determined by UPLC using Acquity chromatograph (Waters). For CD measurements the peptide was dissolved in deionized water at the concentration of 0.35 mg/ml.

\subsection{Circular dichroism (CD) spectroscopy}
CD measurements were performed with the Chirascan CD spectrometer (Applied Photophysics). A 1.0 mm path-length cell was used. CD scans were taken from 190 to 260 nm, with a 1 nm bandwidth and a 1 nm step resolution, three scans were added and averaged for each spectra. In order to estimate secondary structure content averaged CD spectra were analysed using CDNN program\cite{cdnn}. To obtain a melting curve of the peptide structure CD measurements were conducted at different temperatures ranging from $5\; ^{\circ}$C to $90\; ^{\circ}$C with a step of $5\; ^{\circ}$C.

\section{Results and discussion}
\subsection{Peptide folding}
Folding dynamics of the peptide were obtained as a result of the molecular dynamics simulation starting from the linear conformation. We investigated a \emph{root mean square deviation} (RMSD) of the C$^{\alpha}$ atoms that is a measure of distance between different spatial structures. Time evolution of the RMSD of the peptide with it's initial conformation is shown in fig.1. As expected, fast process of structure formation  occurred in the first 30 ns of trajectory that is expressed in the large change of RMSD from zero to roughly 1.8 nm.  After 30 ns RMSD is nearly constant with fluctuation about 0.25 nm, which is the typical value for the thermal vibrations.  A  stable structure appears after the 90 ns and remains constant for subsequent calculations. 
\begin{figure}[h]
\begin{minipage}[t]{0.45\textwidth}
\includegraphics[width=\textwidth]{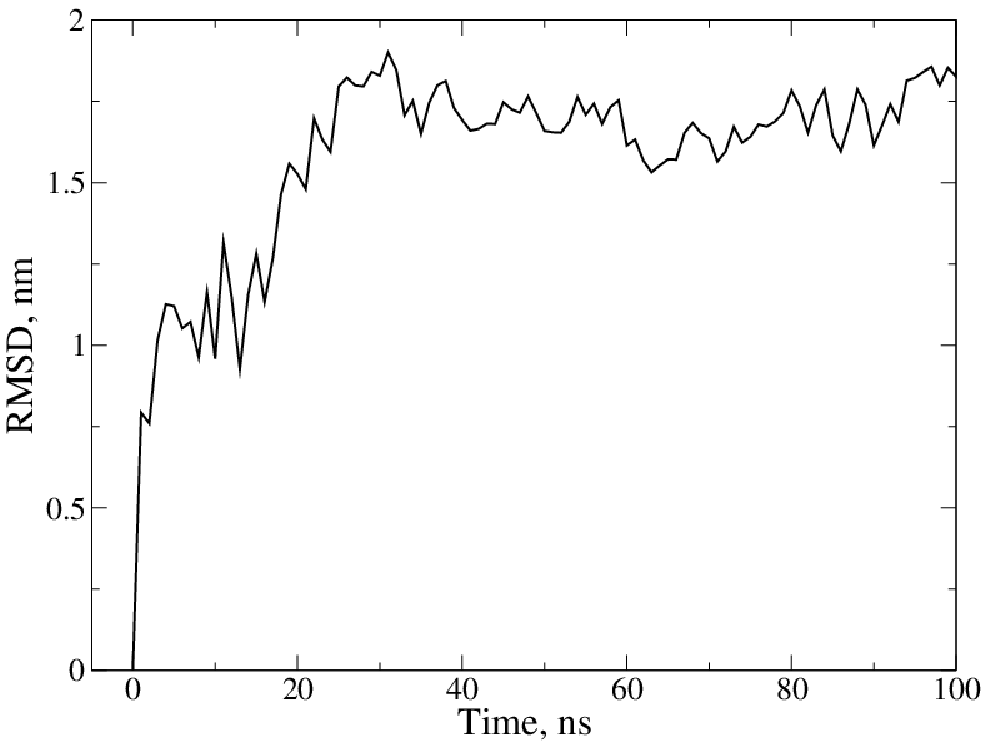}
\caption{Process of the peptide folding. Root mean square deviation (RMSD) of the peptide C$^{\alpha}$ atom positions during the first simulation.}
\end{minipage} \hspace{1cm}
\begin{minipage}[t]{0.45\textwidth}
\includegraphics[width=\textwidth]{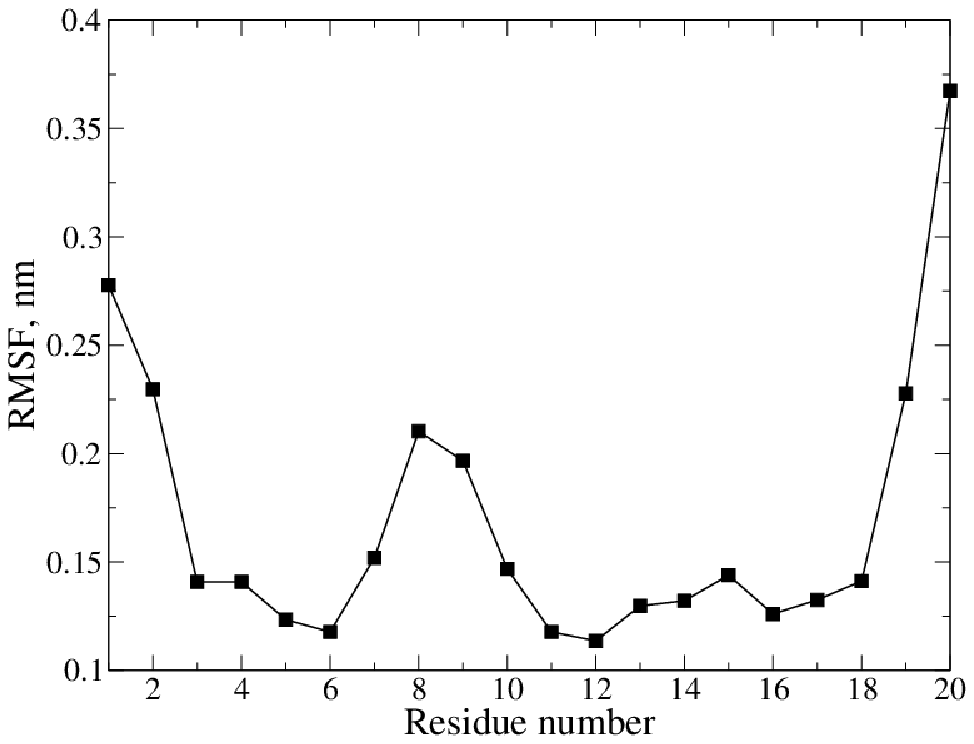}
\caption{Thermal stability of the peptide structure. Root mean square fluctuation (RMSF) of the peptide C$^{\alpha}$ atom positions during the second simulation.}
\end{minipage}
\end{figure} \\
The \emph{root mean square fluctuation} (RMSF) of the C$^{\alpha}$ atoms is a measure of thermal vibrations and rigidity of the peptide structure. RMSF of the D51 calculated from the second simulation is shown on fig.2. Residues 3--6 and 11--18 have the RMSF values below 0.15~nm, so it can be assumed that they belong to the relatively rigid structure motif such as $\beta$-sheet.  Obviously, the C- and N-terminus of the peptide and also a loop between the two expected $\beta$-strands have a large mobility while the atoms in beta-sheets are rather stationary.   

\subsection{Spatial structure and hydrophobicity}
The final peptide conformation after the second simulation was analysed using DSSP program\cite{dssp}. The results of the analysis of hydrogen-bonding patterns and backbone torsion angles $\phi$, $\psi$ and corresponding secondary structure elements are summarized in Table 1.
\setlength{\tabcolsep}{4pt}
\begin{table}[h]
\caption{Secondary structure assignment for the final peptide conformation: ``E'' -- residue belongs to $\beta$-strand element, ``S'' -- turn, ``--'' -- random coil (unstructured).}
\begin{center}
\begin{tabular}{*{21}{l}}
\br
Residue number &1&2&3&4&5&6&7&8&9&10&11&12&13&14&15&16&17&18&19&20\\
\mr
Amino acid &F&L&F&R&V&A&S&K&V&F&P&A&L&I&G&K&F&K&K&K\\
Secondary structure &--&--&--&E&E&E&E&S&S&S&E&E&E&E&S&--&--&--&--&--\\
\br
\end{tabular}
\end{center}
\end{table}\\
It should be mentioned that the secondary structure elements which were found with DSSP algorithm match with rigid regions having lowest RMSF values (fig.2) perfectly. This match is due to hydrogen bonding that occurs between secondary structure elements ($\beta$-strands) and reduces their thermal mobility. 

Schematic representation of the peptide structure is shown in fig.3(a). Secondary structure elements -- antiparallel $\beta$-strands are drawn as arrows.
\begin{figure}[h]
\includegraphics[width=0.5\textwidth]{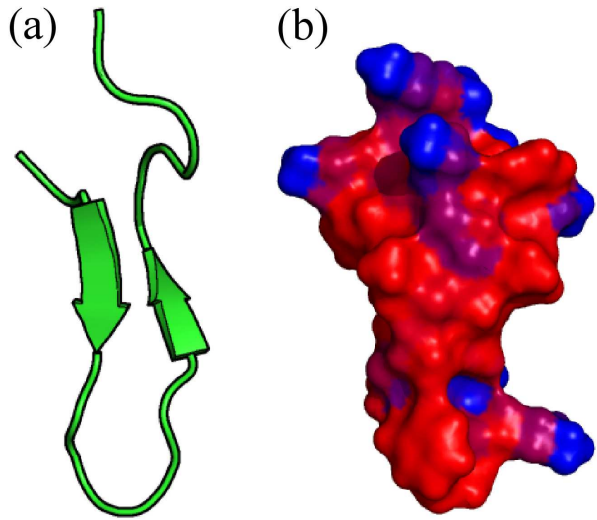} \hspace{1cm}
\begin{minipage}[b]{0.42\textwidth}
\caption{Peptide structure obtained from MD simulations: \\ (a) -- schematic ribbon representation; \\ (b) -- molecular surface representation, red color corresponds to hydrophobic and blue to hydrophilic regions.}
\end{minipage}
\end{figure}

Hydrophobic and hydrophilic (amphiphilic) properties are critical for antimicrobial peptides and their ability to bind to bacterial membranes. The surface of the D51 molecule constructed with PLATINUM program\cite{platinum} using molecular hydrophobicity potential approach is represented in fig.3(b). Blue color corresponds to hydrophilic and red to hydrophobic surface. The most hydrophilic regions correspond to positively charged residues. The highest affinity to the bacterial membrane is usually achieved when spacing between hydrophilic and hydrophobic regions is maximal. In our case the optimal cooperation with the membrane may be interfered by a small hydrophilic area in the main hydrophobic region. This area if formed by polar side chains of S$_{7}$ and K$_{8}$ amino acids .

\subsection{Structure evaluation by CD spectroscopy}
In order to evaluate the secondary structure content of the peptide and explore it's thermal stability a number of CD spectra were measured at different temperatures. Six of the obtained curves presented in the fig.4 show gradual variation of spectra shape with increasing temperature.
\begin{figure}[h]
\begin{minipage}[t]{0.45\textwidth}
\includegraphics[width=\textwidth]{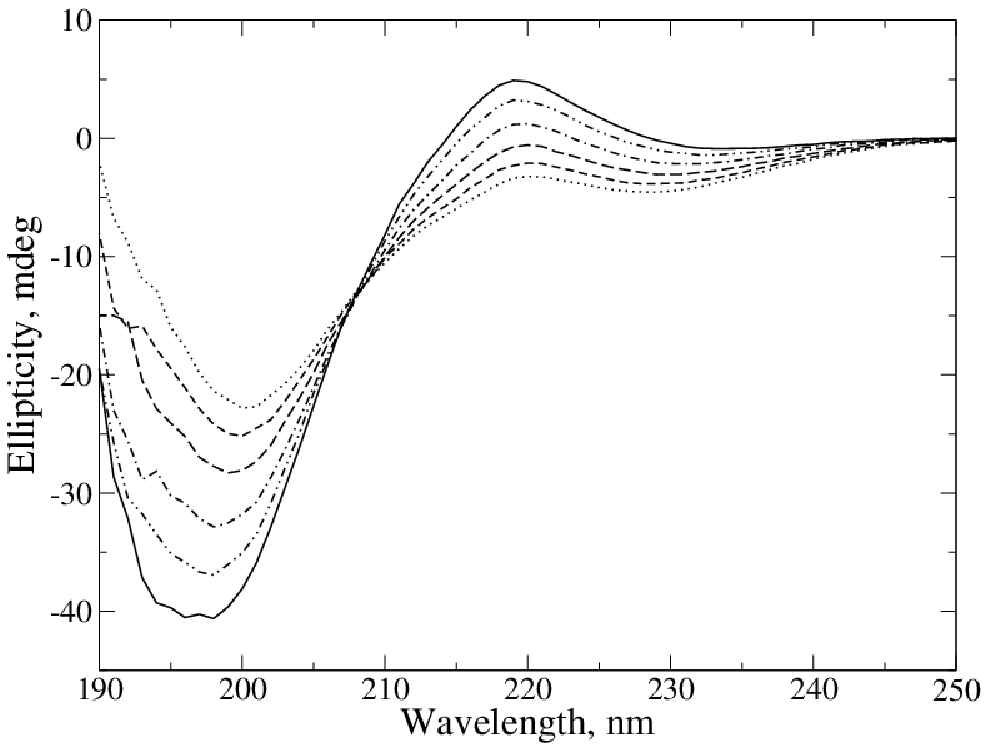}
\caption{CD spectra at different temperatures: $\full$~5$^{\circ}$C, $\dashddot$~20$^{\circ}$C, $\chain$~35$^{\circ}$C, $\broken$~50$^{\circ}$C, $\dashed$~65$^{\circ}$C, $\dotted$~80$^{\circ}$C.}
\end{minipage} \hspace{1cm}
\begin{minipage}[t]{0.45\textwidth}
\includegraphics[width=\textwidth]{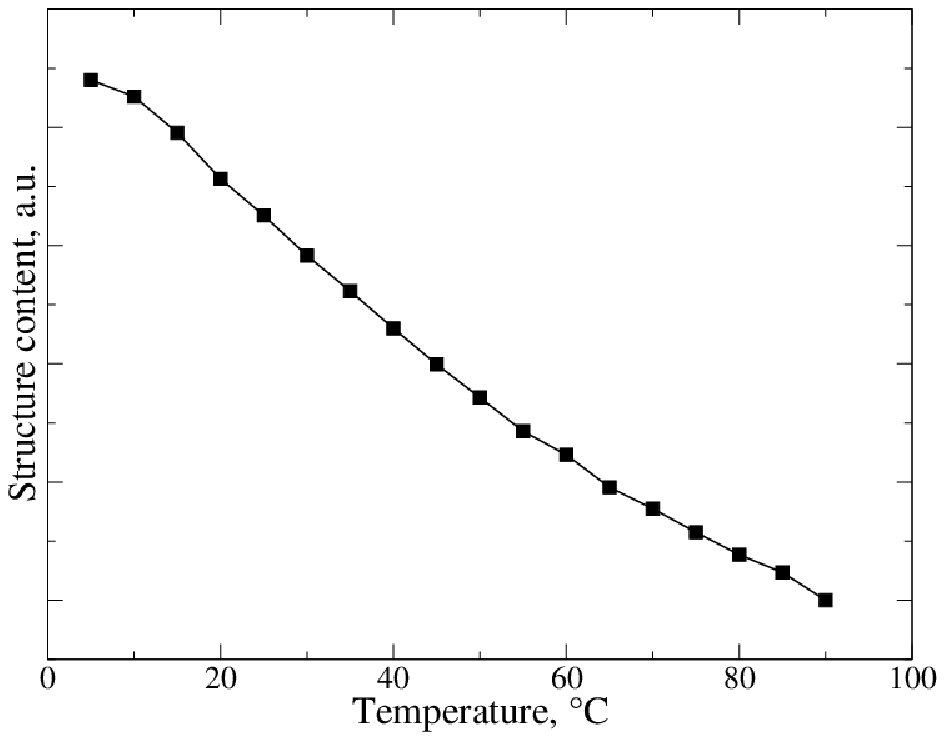}
\caption{Melting curve of the peptide, measured at 220 nm wavelength.}
\end{minipage}
\end{figure}
Thermal stability of the peptide was assessed by construction of melting curve using the values of ellipticity at 220 nm for different temperatures. Melting curve of the peptide (fig.5) demonstrates nearly linear decrease of structure or, in other words, unfolding of the peptide with increasing temperature. There is no characteristic temperature or evidence of saturation in the studied range of temperatures. Such a behaviour, meaning very broad range of temperatures where melting occurs, is typical for relatively small biopolymers\cite{finkel}.

The secondary structure content was determined from the CD measurement with the lowest temperature (5 $^{\circ}$С) and results are listed in Table 2. 
\setlength{\tabcolsep}{6pt}
\begin{table}[h]
\caption{Secondary structure content, \%.}
\begin{center}
\begin{tabular}{*{6}{l}}
\br
Structure element & $\alpha$-helix & Parallel $\beta$-sheet & Antiparallel $\beta$-sheet & Turn & Random coil \\
\mr
MD simulation & 0 & 0 & 40 & 20 & 40 \\
CD measurements & 11 & 9 & 36 & 15 & 29 \\
\br
\end{tabular}
\end{center}
\end{table}
As can be seen from Table 2 there is a rather good agreement in secondary structure content obtained from molecular dynamics simulation and by CD spectroscopy especially for antiparallel $\beta$-sheet. CD measurements also show presence of minor $\alpha$-helix and parallel $\beta$-sheet content that is different to MD results. We assume that this minor components can be neglected because their content is comparable to CD method accuracy.  

\subsection{Hypothesis on the peptide interaction with bacterial membranes}
The mechanism of action of cationic AMPs is still not well understood but it is widely accepted that all antimicrobial peptides interact with bacterial membranes\cite{review}. It has been shown that cationic peptides have a higher affinity to LPS in the membrane of the Gram-negative bacteria than native divalent cations such as Mg$^{2+}$ and Ca$^{2+}$, so peptides displace these cations from the negatively charged LPS leading to a local disturbance in the outer membrane\cite{structure-activity}. This facilitates the formation of destabilized areas through which the peptide with it's hydrophobic part integrates into lipid bilayer.
 
Therefore, for the disruption of bacterial membranes the D51 peptide should first bind to the outer membrane with it hydrophilic areas. Presence of the highly hydrophilic positively charged region (on the top of fig.3(b)) formed by three Lys residues at the C-terminus could account for high electrostatic affinity to LPS. However from the spatial structure it can be seen that there is a charged residue in the hydrophobic area, which can impede further penetration of the peptide into lipid bilayer. 

Considering obtained spatial structure and hypothetical mechanism of action of the studied peptide we propose rational modification of it's sequence. One of the most perspective permutation is to move the positively charged K residue to the N-terminus of the peptide: KFLFRVASVFPALIGKFKKK.  This permutation leads to further separation of hydrophobic and hydrophilic regions that possibly can improve peptide activity against Gram-negative bacteria. 

\section{Conclusion}
Rational design of new antimicrobial peptides with high antibacterial activity is an important step on the way to the development of new generation antibiotics against multidrug-resistant bacteria. To date, several methods of rational design of AMPs exist, but all of these methods do not take into account spatial structure of peptides which is essential for their action.

In this work we applied methods of molecular dynamics simulation and circular dichroism spectroscopy to give insights into spatial structure organization and possible mechanism of action of previously designed synthetic antimicrobial peptide D51. Structural properties such as secondary structure content and hydrophobicity potential distribution on the molecular surface were determined by molecular dynamics study.

The peptide was chemically synthesized and it's secondary structure was investigated by CD spectrometry, confirming our theoretical results. Considering a possible mechanism of the peptide interaction with bacterial membranes a rational modification of it's structure was proposed.

\section*{Acknowledgments} 
Authors wish to acknowledge technical support by the SPC facility at EMBL Hamburg. Joint Supercomputer Center of the Russian Academy of Sciences (JSCC RAS) is acknowledged for provided computational facilities. This work was financially supported by Presidium of the
Russian Academy of Sciences as a part of the ``Development of a technology for the design of new antimicrobial peptides with the use of methods of bioinformatics and molecular modeling'' project.

\section*{References}
\providecommand{\newblock}{}

\end{document}